\documentclass[prl,aps,twocolumn,amsfonts,showpacs,footinbib,preprintnumbers,superscriptaddress]{revtex4-1}

\usepackage{graphicx}
\usepackage{xcolor}
\usepackage{rotating}
\usepackage{amsmath,amssymb,graphics}

\usepackage[colorlinks=true, urlcolor=violet, linkcolor=blue, citecolor=red, hyperindex=true, linktocpage=true]{hyperref}

\newcommand{\nn}{\nonumber\\}

\def\CE{\mathcal{E}}

\def\CN{\mathcal{N}}
\def\CO{\mathcal{O}}
\def\CP{\mathcal{P}}
\def\CS{\mathcal{S}}

\def\lgb{\lambda_{\scriptscriptstyle GB}}

\begin{document}

\title{Coupling constant corrections in a holographic model of heavy ion collisions}

\author{Sa\v{s}o Grozdanov}
\affiliation{Instituut-Lorentz for Theoretical Physics, Leiden University, Niels Bohrweg 2, Leiden 2333 CA, The Netherlands}

\author{Wilke van der Schee}
\affiliation{Center for Theoretical Physics, MIT, Cambridge, MA 02139, USA}

\preprint{MIT-CTP/4850}


\begin{abstract}
We initiate a holographic study of coupling-dependent heavy ion collisions by analysing for the first time the effects of leading-order, inverse coupling constant corrections. In the dual description, this amounts to colliding gravitational shock waves in a theory with curvature-squared terms. We find that at intermediate coupling, nuclei experience less stopping and have more energy deposited near the lightcone. When the decreased coupling results in an 80\% larger shear viscosity, the time at which hydrodynamics becomes a good description of the plasma created from high energy collisions increases by 25\%. The hydrodynamic phase of the evolution starts with a wider rapidity profile and smaller entropy.  
\end{abstract}

\maketitle

\noindent
{\bf 1. Introduction.---}Relativistic collisions of heavy ions at RHIC and LHC result in formation of a strongly interacting state of matter known as the quark-gluon plasma (QGP). While these experiments provide an invaluable window into properties of quantum chromodynamics (QCD), our theoretical understanding of QGP in QCD remains far from complete. In recent years, gauge-gravity duality (holography) has enabled theoretical studies of certain, usually supersymmetric, classes of large-$N$ field theories, which are most readily performed at infinitely strong ('t Hooft) coupling $\lambda$. As a result of those advances, many properties of QGP previously conceived as impenetrably complex, such as its collective far-from-equilibrium behaviour, can now be analysed using numerical general relativity techniques. At infinite coupling, heavy ion collisions have been successfully modelled by (dual) collisions of gravitational shock waves in Einstein bulk theory with an extra dimension and a negative cosmological constant \cite{Chesler:2010bi,Grumiller:2008va,Casalderrey-Solana:2013aba,Casalderrey-Solana:2013sxa,Chesler:2015wra} (see \cite{jorge-book,Chesler:2015lsa,Heller:2016gbp} for reviews.). 

We face several challenges in connecting holography with experimental studies of QGP, which typically occur at the intermediate coupling strength. Most formidable among them is establishing a bulk dual to non-supersymmetric Yang-Mills theory with any (small) number of colours and fundamental matter. From the point of view of presently understood holography, even computing $1/N$ corrections around infinite $N$ requires inclusion of quantum gravity corrections (perturbative topological $g_s$ corrections in string theory) (see e.g. \cite{Denef:2009yy,Denef:2009kn,Arnold:2016dbb}). Including coupling constant corrections in a perturbative ($1/\lambda$) series around infinite coupling is easier and requires one to find an $\alpha'$-corrected higher-derivative supergravity action. To leading order, such coupling constant corrections have been computed for thermodynamics \cite{Gubser:1998nz}, hydrodynamics \cite{Kats:2007mq,Brigante:2007nu,Brigante:2008gz,Buchel:2008ae,Grozdanov:2014kva,Grozdanov:2015asa}, thermalisation and higher-frequency spectrum at linear \cite{Stricker:2013lma,Waeber:2015oka,Grozdanov:2016vgg} and non-linear level \cite{Andrade:2016aaa}. Ref. \cite{Grozdanov:2016vgg} further showed that simple leading-order higher derivative corrections to the bulk action reproduce a variety of coupling constant dependent phenomena, including the approach to kinetic theory regime and breakdown of hydrodynamics above a coupling-dependent critical momentum.

In this work, we describe the first dynamical (real-time) collision with coupling constant dependence. The dynamical nature allows us to see how the system evolves towards a hydrodynamic plasma, how the energy distributes itself, and to study the entropy production during the collision. What we will demonstrate is that as the coupling constant is decreased, the nuclei experience less stopping with more energy deposited on the lightcone and have a flatter distribution of energy in the plasma. The time until the effective hydrodynamic description becomes applicable (hydrodynamisation time, $t_{\rm hyd}$) is increased and less total entropy is produced.

\begin{figure*}[t]
\begin{center}
\includegraphics[width=9.0cm]{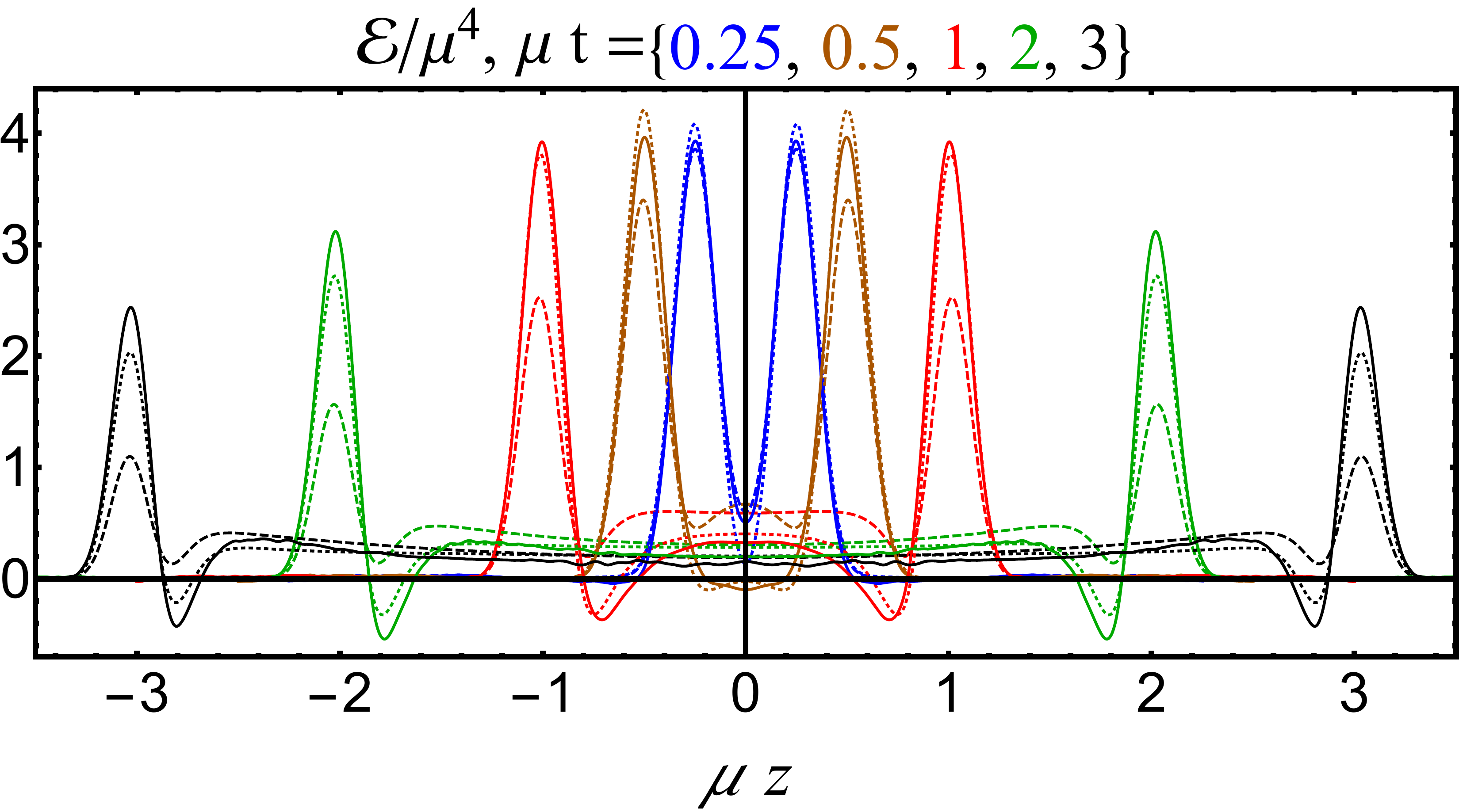}\,\,
\includegraphics[width=7.6cm]{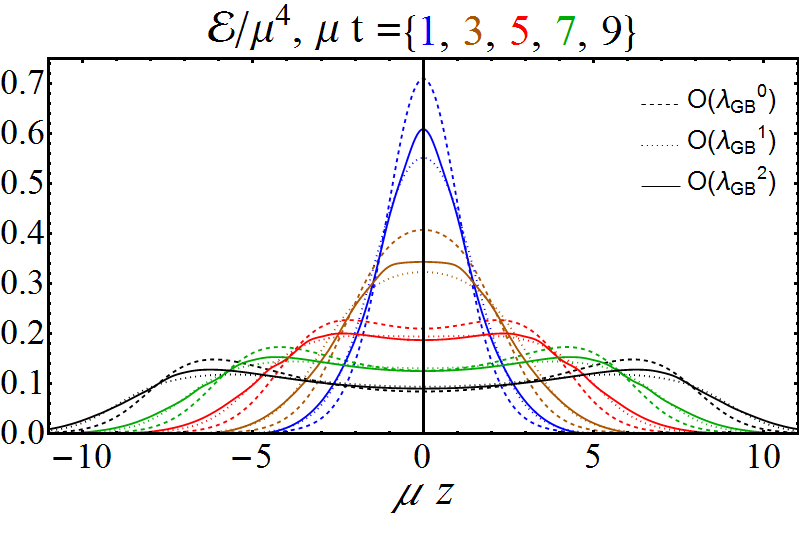}
\end{center}
\caption{
Energy density along the longitudinal coordinate $z$ at different times for narrow (left) and wide shock collisions (right). We present results for infinitely coupled (dashed), first-order corrected (dotted) and second-order corrected (solid) collisions at $\lgb = -0.2$. At intermediate coupling, we observe increased energy density near the lightcones (peaks), which signals less stopping. In the plasma (between peaks), the energy density is flatter, which is caused by a smaller longitudinal pressure due to the larger viscosity.
}
\label{fig:snapshots}
\end{figure*}

\noindent
{\bf 2. Curvature-squared theories.---}We will restrict our attention to the simplest, leading-order class of perturbative (in $\alpha'$) higher-derivative corrections and study curvature-squared theories. Such effective supergravity actions of massless modes, which are known to generically arise from string theory, can be found by either computing loop corrections to the world-sheet beta function \cite{Callan:1986bc,Grisaru:1986vi} or by guessing the right action that could result in scattering amplitudes computed from string theory \cite{Gross:1986mw,Gross:1986iv,Freeman:1986zh}. It is important to note that type IIB string theory compactified on $S^5$, dual to $\mathcal{N} = 4$ SYM theory, is special from the point of view that all $\alpha'$ and $\alpha'^2$ corrections vanish and the leading-order corrections are proportional to $\alpha'^3 R^4$ with $\alpha' \propto L^2  / \lambda^{1/2}$. We adopt the view that it is plausible that in more realistic theories, such as in a putative dual to QCD, the leading order corrections would enter at order $\alpha'$, which is why we restrict to $R^2$ theories. Nevertheless, we wish to stress that a precise theory dual to these theories is unknown, even though the holographic framework allows to compute field theory quantities, such as the expectation values of dual operators.

At leading order in $\alpha'$, the most general curvature-squared action can be written as the Einstein-Gauss-Bonnet theory \footnote{Any perturbative $R^2$ theory can be conveniently transformed into the Einstein-Gauss-Bonnet theory (with second-order equations of motion). For details, see \cite{Brigante:2007nu,Grozdanov:2014kva,Grozdanov:2016vgg}.}
\begin{align}
\label{GBaction}
S_{GB} &= \frac{1}{2\kappa_5^2} \int d^5 x \sqrt{-g} \biggr[ R  - 2 \Lambda \nn
&+ \frac{\lgb}{2} L^2 \left( R^2 - 4 R_{\mu\nu} R^{\mu\nu} + R_{\mu\nu\rho\sigma} R^{\mu\nu\rho\sigma} \right) \biggr],
\end{align}
where $\lgb\propto\alpha'$, which we will treat perturbatively \footnote{Recently, \cite{Camanho:2014apa} argued that \eqref{GBaction} violates causality unless $|\lgb| / L^2 \ll 1$ (see however \cite{Papallo:2015rna,Andrade:2016yzc}). Since we work perturbatively in $\lgb$, such restrictions should not affect our findings.}. The negative cosmological constant $\Lambda = - 6 / L^2$ sets the anti-de Sitter scale, for which, to first order, we choose $L\equiv L_0 + \lgb L_1 =1+\lgb/2 $ \footnote{An advantage of this choice is that the non-normalisable mode of the metric does not receive corrections.}.

As discussed in \cite{Grozdanov:2016vgg,Grozdanov:2016fkt}, Einstein-Gauss-Bonnet theory qualitatively reproduces the departure from infinitely coupled physics towards weaker coupling when $\lgb < 0$, including a larger shear viscosity ($\eta/s=\frac{1}{4\pi}(1-4\lgb)$) \cite{Brigante:2007nu}. Despite the perturbative nature of our calculation in $\lgb$, for clarity, we will present results for $\lgb = -0.2$ and demonstrate convergence by computing the energy profiles up to $\CO(\lgb^2)$. In $\CN=4$ theory, the value of the 't Hooft coupling that increases $\eta/s$ by $80\%$ is $\lambda \approx \CO(10)$.

\begin{figure}
\begin{center}
\includegraphics[width=8.5cm]{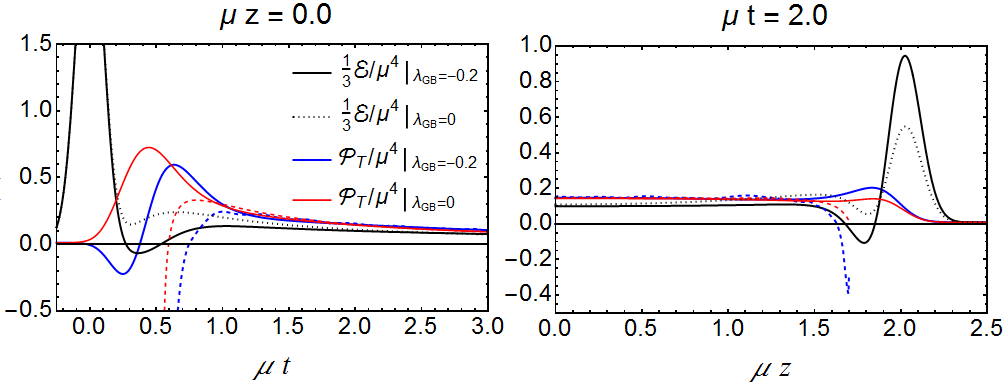}
\includegraphics[width=8.5cm]{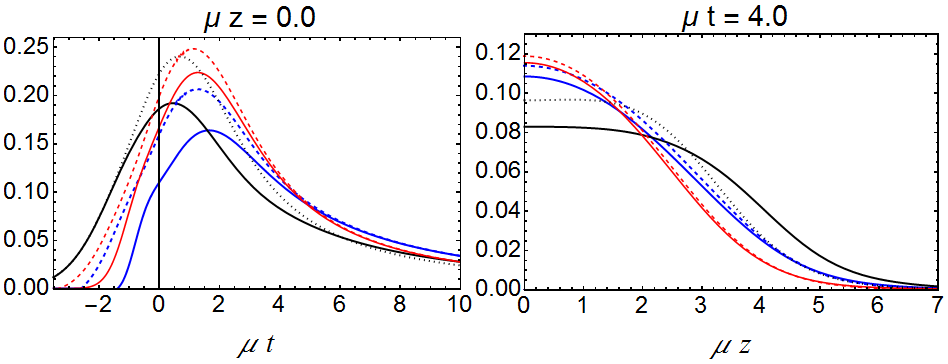}
\end{center}
\caption{Plots of infinitely coupled $\CE/\mu^4$ (black, dotted) and $\CP_T / \mu^4$ (blue, solid), and $\CE/\mu^4$ (black, solid) and $\CP_T / \mu^4$ (red, solid) at intermediate coupling, as well as hydrodynamic predictions for pressures (dotted) as functions of time $t$ and longitudinal coordinate $z$. Top plots represent narrow and bottom plots wide shocks. For narrow shocks, hydrodynamics breaks down near the lightcone.}
\label{fig:hydro}
\end{figure}

We work with a metric ansatz given in the characteristic formulation of \cite{Bondi1960Gravitational,Chesler:2010bi} by
\begin{align}\label{metric}
ds^2 = \! - A dt^2 \!+ 2 dt \left( dr +  F dz \right) \!+ S^2 \! \left( e^B d {\bf x}_{\perp}^2\! + e^{-2B} dz^2 \right) ,
\end{align}
where the functions $A$, $B$, $F$ and $S$ all depend on the Eddington-Finkelstein time $t$, the (holographic) radial coordinate $r$ and the longitudinal coordinate $z$.
Perturbatively (in $\lgb$), we write $A = A_0 + \lgb A_1$ and similarly for $B$, $F$ and $S$. After expanding the equations of motion derived from \eqref{GBaction} to first order in $\lgb$, we have to solve two sets of differential equations. First, the standard nested set of ordinary differential equations (ODEs) for $A_0$, $B_0$, $F_0$ and $S_0$ \cite{Chesler:2010bi}, followed by an almost identical nested set of non-homogeneous ODEs for $A_1$, $B_1$, $F_1$ and $S_1$. The non-homogeneous terms depend on the numerical solution of the zeroth-order functions. 

\begin{figure*}[t]
\begin{center}
\includegraphics[width=6.2cm]{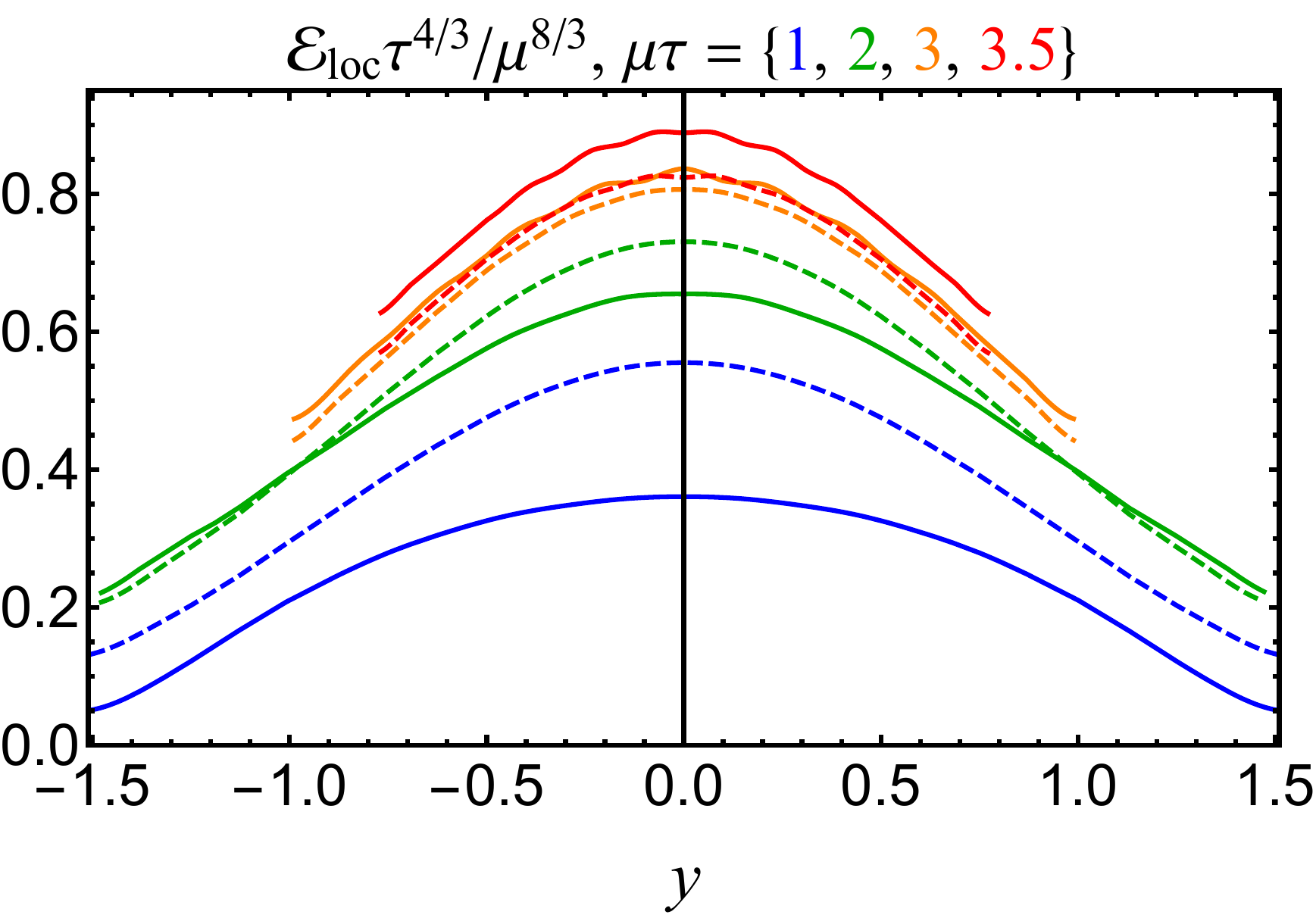}\quad
\includegraphics[width=6.2cm]{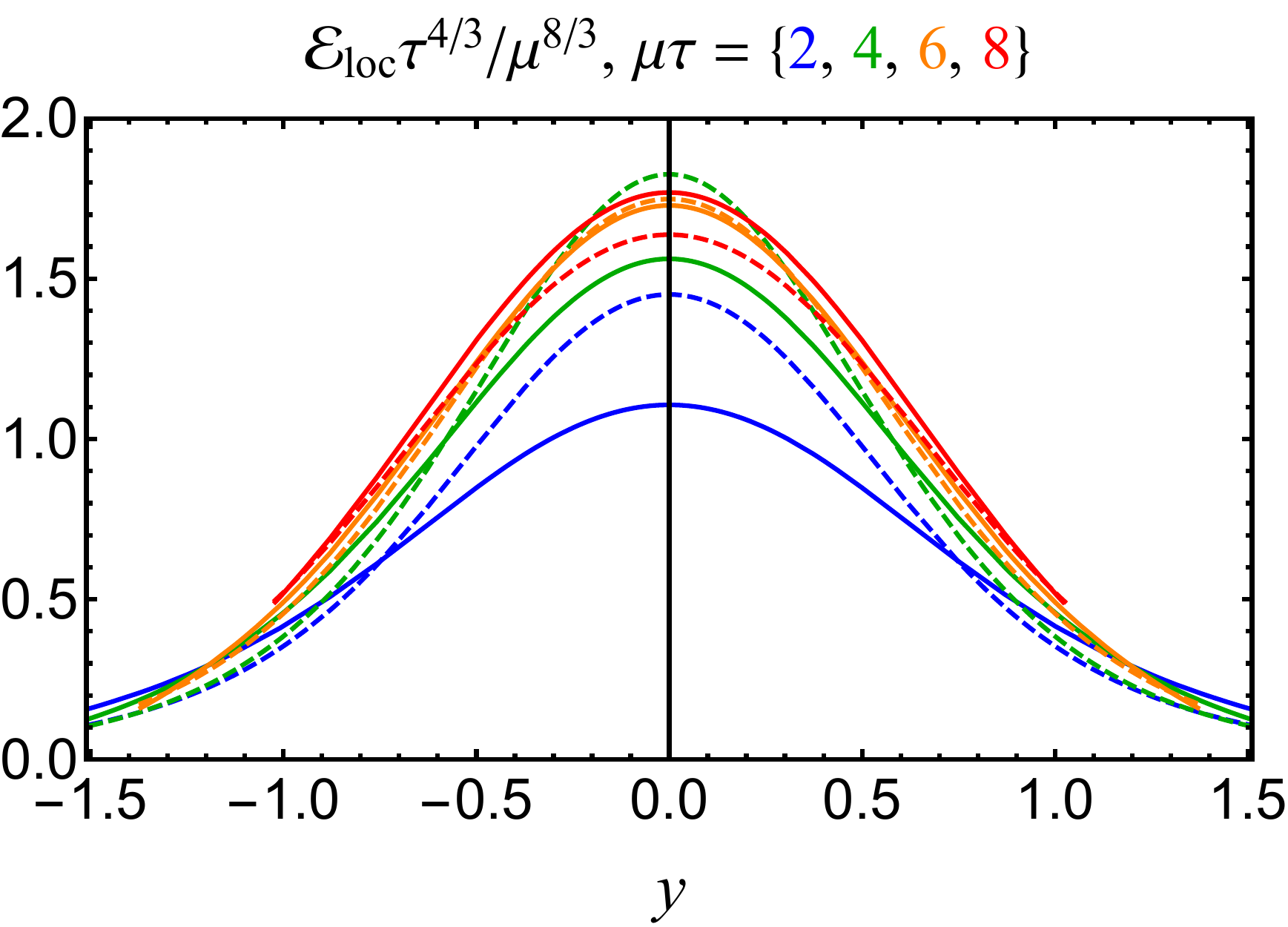}
\end{center}
\caption{Rapidity distributions (multiplied with $\tau^{4/3}$ to compensate for the expansion of the plasma) for narrow (left) and wide (right) shock collisions at $\lgb=-0.2$ (solid) and $\lgb=0$ (dashed). The distributions start out wider and smaller, but become of comparable width and amplitude due to the 80\%  larger viscosity.}
\label{fig:rapidity}
\end{figure*}

Once the solution is found, it is well-known how to obtain the dual (conformal) holographic stress-energy tensor (see e.g. \cite{Brihaye:2008kh,Grozdanov:2015asa}). We introduce a rescaled energy density
\begin{align}\label{Energy}
\mathcal{E} \equiv  \frac{\kappa_5^2}{2 L_0^3} T_{tt}  = -\frac{3}{4} \left(a_{4,0} + \lgb (a_{4,1}-2 a_{4,0})\right),
\end{align} 
where $a_{4,0}$ and $a_{4,1}$ are the normalisable modes of $A_0$ and $A_1$, respectively, with analogous formulas for the pressures. At $\lgb=0$, the prefactor equals $ 2 \pi^2 / N_c^2$ for the dual $\CN=4$ SYM. For duals of the Einstein-Gauss-Bonnet theory, such a relation is unknown.

We are interested in studying the collision of planar sheets of energy, dual to the collision of gravitational shock waves. For our choice of $L$, the single shock wave metric in Fefferman-Graham coordinates \cite{Janik:2005zt,Chesler:2010bi} continues to be an exact solution of the equations of motion. These sheets of energy are characterised by a single non-zero component of the stress-energy tensor:
\begin{align}\label{FGmetric}
T_{\pm \pm }(z_\pm) =  \frac{\kappa_5^2}{2 L_0^3}  \, \frac{\mu^3}{\sqrt{2\pi w^2}}\, e^{-z_\pm^2/2 w^2}, 
\end{align}
with $z_\pm = t\pm z$ and $w$ the width of the sheets. The sign in $z_\pm$ depends on the direction of motion of the shock. We can easily find a metric, such that the rescaled energy per transverse energy, $\mu^3$, does not depend on $\lgb$ (this implies $a_{4,1} = 2 a_{4,0}$, initially). These initial conditions can then be translated to Eddington-Finkelstein coordinates with the standard method explained in \cite{Chesler:2013lia,vanderSchee:2014qwa}.

\noindent
{\bf 3. Heavy ion collisions at finite coupling.---}We present results for narrow and wide shocks, with $\mu w = 0.1$ and $\mu w = 1.5$ \footnote{For numerical stability, our simulations include a regulator energy density of less than 1\% of the peak energy density. Further technical specifications, including the \emph{Mathematica} code used for the simulations, can be found on \href{http://sites.google.com/site/wilkevanderschee/}{sites.google.com/site/wilkevanderschee/}.}. Due to the Lorentz contraction at high energies, it is possible to think of narrow and wide shocks as of high- and low-energy heavy ion collisions, respectively. In Fig. \ref{fig:snapshots}, we first present snapshots of the energy density profiles after collisions to $\CO(\lgb^2)$ \footnote{The calculation to second order in $\lgb$ is entirely analogous to the first-order calculation described above. We expand $A = A_0 + \lgb A_1 +\lgb^2 A_2$ and set $L =1+\lgb/2 + 7 \lgb^2 / 8$, which gives $\mathcal{E}  = -\frac{3}{4} \left(a_{4,0} + \lgb (a_{4,1}-2 a_{4,0}) +\lgb^2 ( a_{4,2} - 2 a_{4,1} - 2 a_{4,0}  )   \right)$. Our initial conditions thus require setting $a_{4,2} = 2 a_{4,1} + 2 a_{4,0}$.}. 
In \cite{Casalderrey-Solana:2013aba,Grumiller:2008va}, it was noticed that narrow shocks exhibit a transparent regime on the lightcone as they pass through each other. At weaker coupling, we find that this effect of transparency is greatly enhanced. More precisely, for $\lgb=-0.2$, to leading-order in $\lgb$, the maximum energy density on the lightcone is $88\%$ higher at the end of our simulation then for $\lgb = 0$. The energy deposited in the plasma is consequently smaller and we find its distribution to be flatter---a point to which we will return shortly. 
The difference between first and second order corrected results is small, but it is comforting to see that the 2nd order result reduces the increase in energy on the lightcone for the 1st order result around time $\mu t=0.4$.

For wide shocks, \cite{Casalderrey-Solana:2013aba} found a curious feature: not only did the energy come to a full stop and explode hydrodynamically, but due to the strong interactions, the energy piled up at mid-rapidity, leading to a maximum energy density of 2.71 times the maximum energy density of the incoming nuclei. Now, at weaker coupling, this effect subsides very rapidly; for $\lgb=-0.2$, the maximum $\CE / \mu^4$ is only $2.17$ times the initial maximum.

\begin{figure}[t]
\begin{center}
\includegraphics[width=4cm]{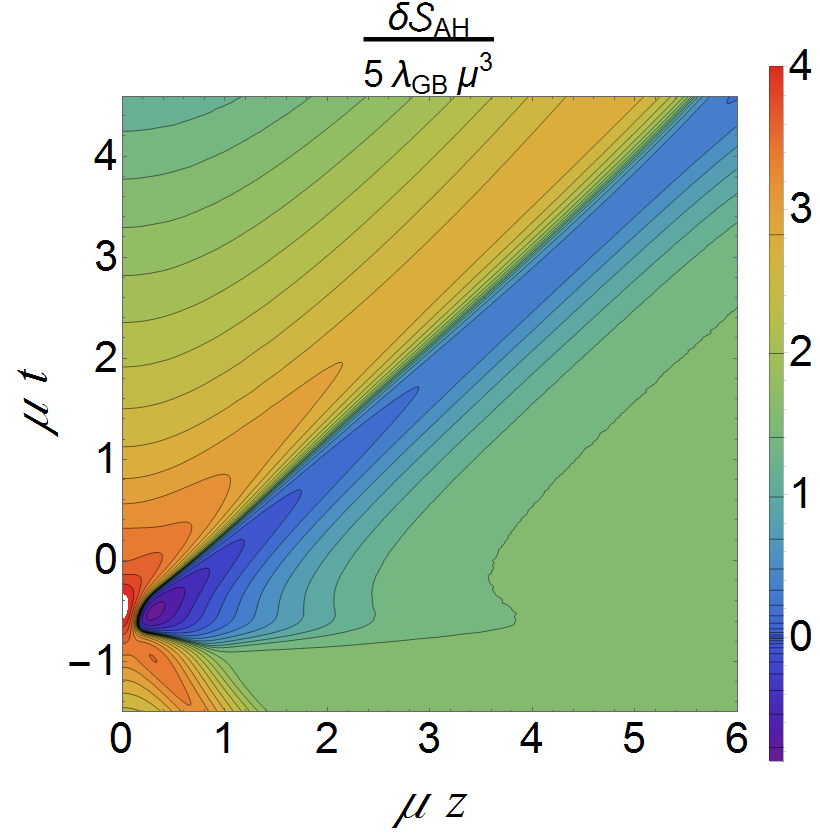}\quad
\includegraphics[width=4cm]{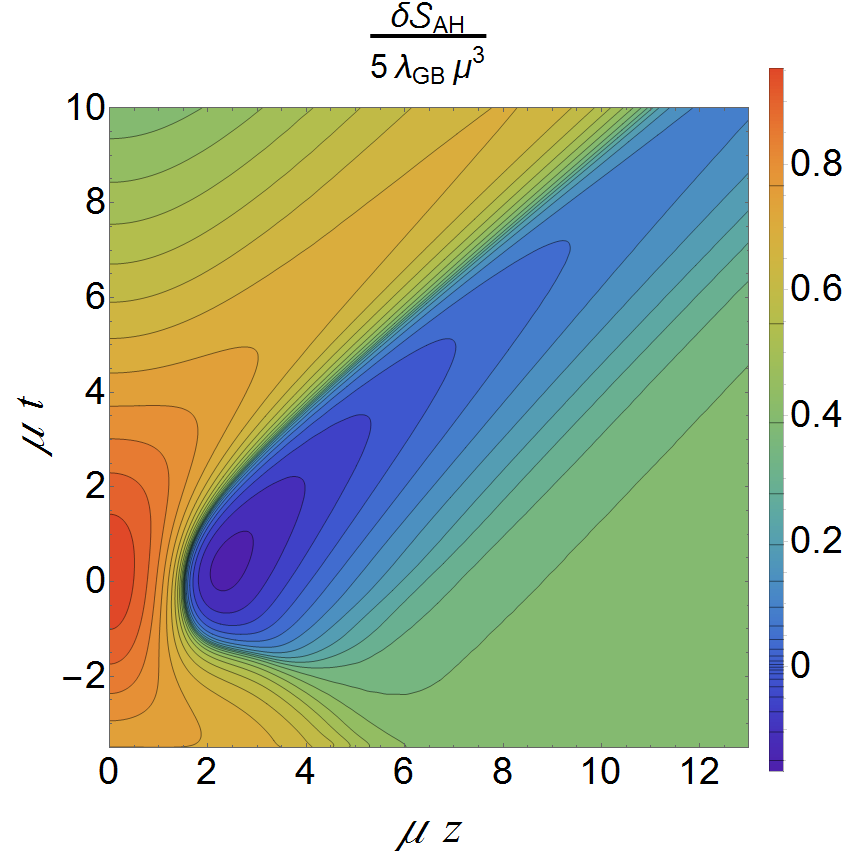}
\end{center}
\caption{Coupling constant correction to entropy density as measured by change in the area of the apparent horizon for narrow and wide shock waves. For negative $\lgb$ the entropy is enhanced at the lightcone, while negative in the plasma. At even later times the entropy correction also becomes positive at mid-rapidity due to the larger viscous entropy production.}
\label{fig:EntropyDifference}
\end{figure}

One of the hallmarks of infinitely strongly coupled collisions is its rapid relaxation towards a hydrodynamic regime (hydrodynamisation). Given some temperature, this occurs within a time of $ t_{\rm hyd} < 0.5 / T$ at mid-rapidity---a result that is independent of the width of the shocks \cite{Casalderrey-Solana:2013aba}. This comparison is usually done by comparing the full far-from-equilibrium pressure with the pressure that would follow from first-order hydrodynamics, given some temperature and fluid velocity. From our simulations, we extract the $\lgb$-dependent temperature, fluid velocity and viscosity. Combined, they allow us to compare the pressure (solid) with pressure computed from hydrodynamics (dashed), as shown in Fig. \ref{fig:hydro}, for both $\lgb=0$ (red) and $\lgb=-0.2$ (blue) (note that the longitudinal pressure follows from $\CE=2\CP_T+\CP_L$).

It is clear that at intermediate coupling, it takes longer for hydrodynamics to become a good description of the evolution of the plasma. We quantify this by obtaining the time when hydrodynamics describes the transverse pressure within 10\%, for small $\lgb$, which we find to be 
$\mu t_{\rm hyd} = \{1.48 - 1.1 \lgb, 1.40 - 19\lgb\} $ for \{narrow, wide\} shocks. By further including the change in temperature, this leads to the hydrodynamisation time in units of the temperature at the time of hydrodynamisation:
\begin{align}
t_{\rm hyd} T_{\rm hyd} = \{0.41 - 0.52\lgb, 0.43 - 6.3 \lgb\},
\end{align}
again for \{narrow, wide\} shocks. For $\lgb=-0.2$, this results in 25\% and 290\% longer hydrodynamisation times. The strong difference between wide and narrow shocks can be explained by realising from Fig. \ref{fig:hydro} that wide shocks approach the hydrodynamic regime much slower due to the continuous inflow of matter. A coupling-dependent perturbation then leads to correspondingly larger hydrodynamisation times. Furthermore, the leading order $t_{\rm hyd} T_{\rm hyd} = 0.43$ for wide shocks is quite sensitive to our criterion for hydrodynamisation
and could be in the range of $t_{\rm hyd} T_{\rm hyd} = 0.0 - 2.0 $ for similar criteria. This would greatly affect the 290\% found at our 10\% criterion. The result for narrow shocks has a very weak dependence on the width and the criterion used, making the increase of $t_{\rm hyd} T_{\rm hyd} $ by 25\% more robust.

To set the initial conditions for the hydrodynamic evolution, it is crucial to obtain the rapidity distribution of the energy deposited in the plasma, which is shown in Fig. \ref{fig:rapidity}. We define rapidity $y$ by $t = \tau \cosh y$ and $z = \tau \sinh y$, where $\tau$ is the proper time. At early times, the finite coupling corrections result in less energy deposited in the plasma, as well as in a wider rapidity profile (at its maximum, the rapidity profile is \{11\%, 23\%\} wider for \{narrow, wide\} shocks). At late times, the increased viscosity at intermediate coupling plays an important role as it decreases the longitudinal pressure. The rapidity profile at $\lgb=0$ grows faster in width and consequently has smaller energy density at mid-rapidity.

\begin{figure}
\begin{center}
\includegraphics[width=4.25cm]{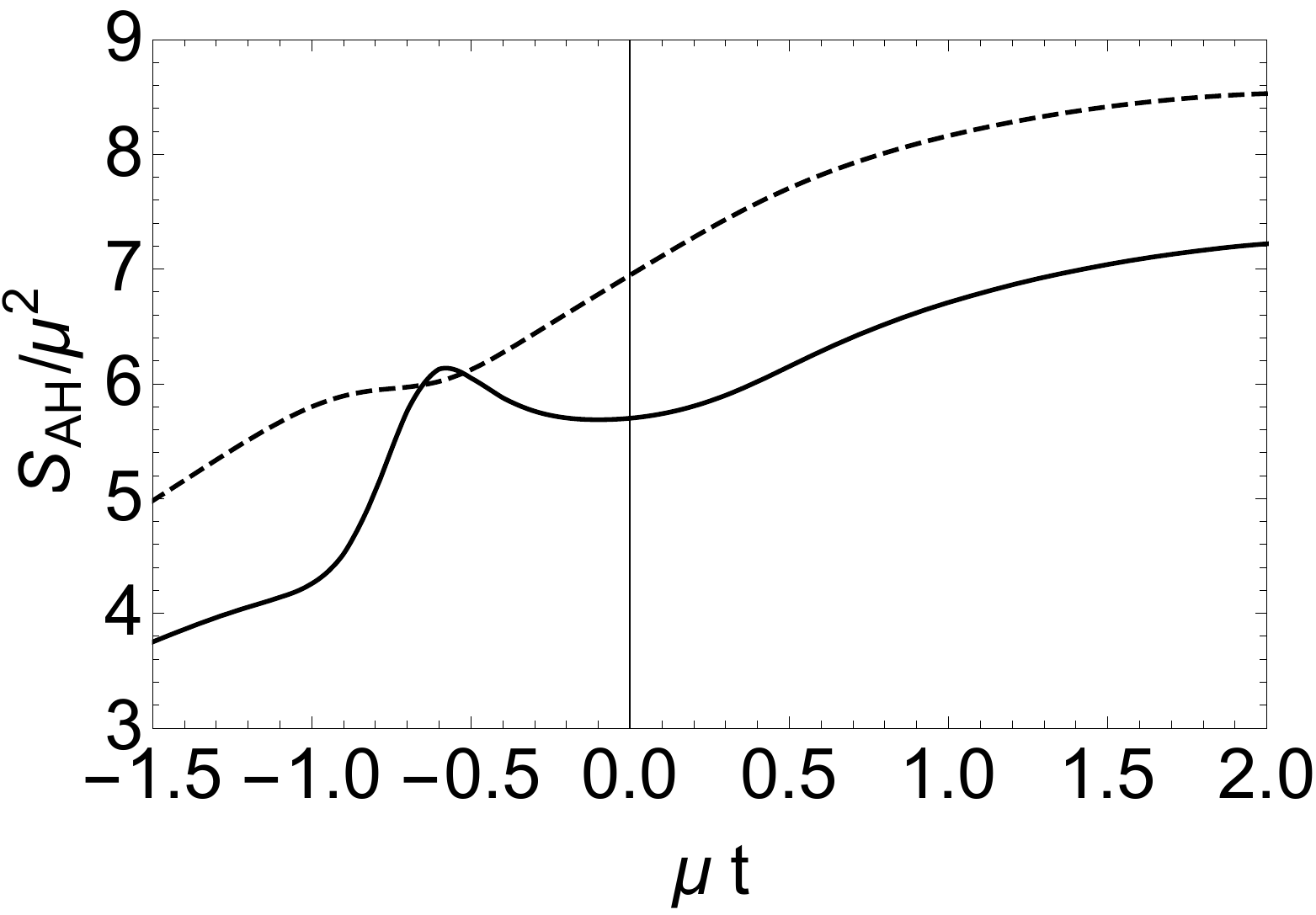}
\includegraphics[width=4.25cm]{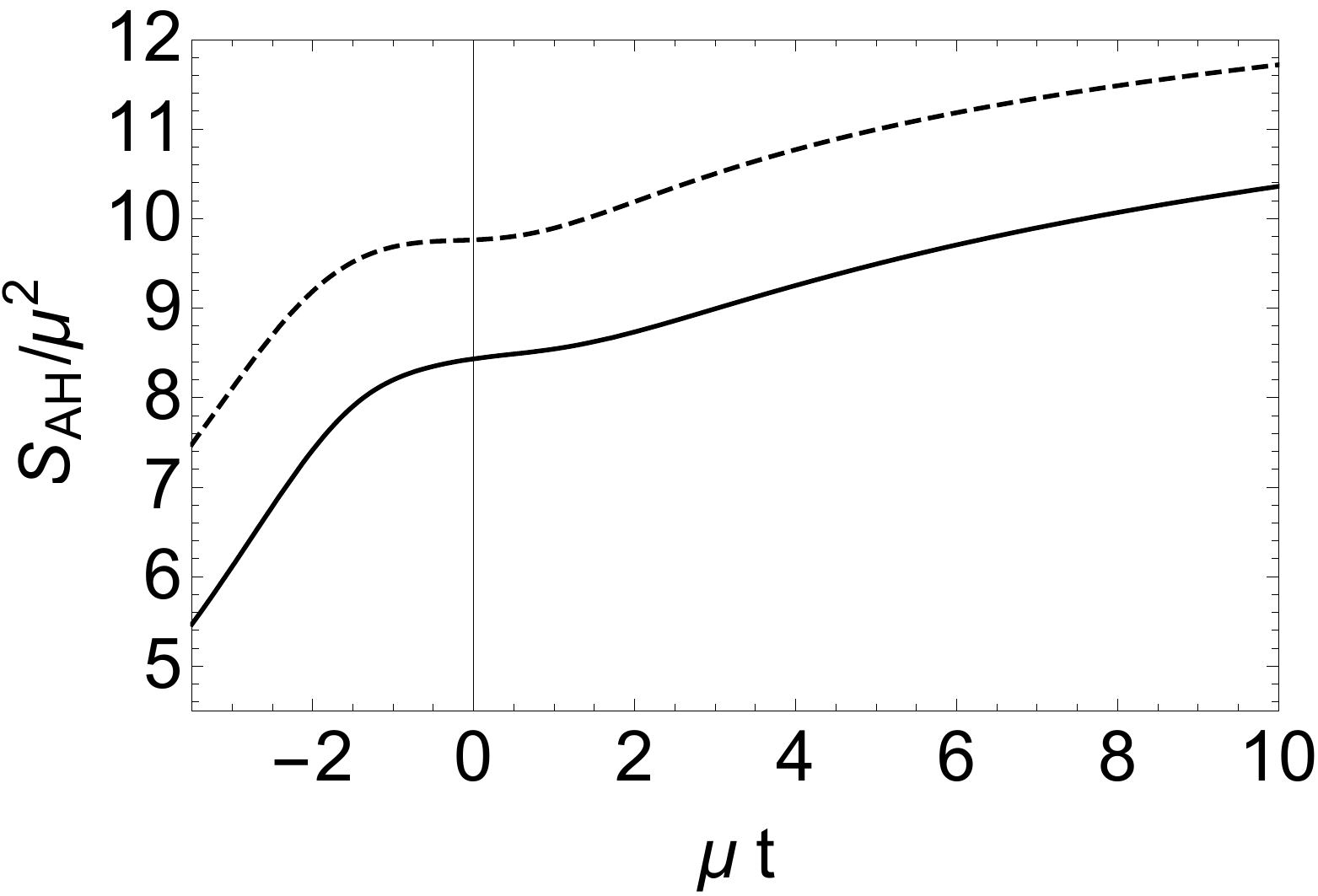}
\end{center}
\caption{Total entropy produced as measured by the apparent horizon area for narrow (left) and wide shocks (right) at $\lgb = -0.2$ (solid) and $\lgb = 0$ (dashed). The decrease in entropy around $\mu t = - 0.5$ is expected to disappear in a simulation with a non-perturbative $\lgb$.}
\label{fig:entropy}
\end{figure}

Another consequence of the increased viscosity can be seen in the entropy production. To demonstrate this, we show both the total entropy per transverse energy as a function of time (Fig. \ref{fig:entropy}) and the difference in the entropy density $\delta s_{AH}$ (Fig. \ref{fig:EntropyDifference}) as measured by the apparent horizon and defined as
\begin{align}
\mathcal{\delta \CS_{AH}} \equiv   \frac{\kappa_5^2}{2 L_0^3}  \delta s_{AH} =3 \pi S_0^2(\lgb S_1+\delta r_{ah} \partial_r S_0).
\end{align}
All quantities are evaluated at the apparent horizon and $\delta r_{ah}$ is the perturbation of its $\lgb=0$ position that depends on $t$ and $z$. It is clear from Fig. \ref{fig:EntropyDifference} that for negative $\lgb$, there is more entropy near the lightcone and less in the plasma---in agreement with Fig. \ref{fig:snapshots}. Overall, this leads to less total entropy, with 27\% and 25\% reduction for narrow and wide shocks at the start of our simulation. Due to larger viscous entropy production, this difference decreases in the hydrodynamic regime and we find 15\% and 12\% reduction for narrow and wide shocks at the end of Fig. \ref{fig:entropy}. The decrease can be partly explained by a reduced number of degrees of freedom, as our choice of $L$ and $\lgb$ implies 15\% fewer degrees of freedom, as measured by $T_{tt}/T^4$. Lastly, we stress that Fig. \ref{fig:entropy} implies that the choice of $\lgb=-0.2$ is likely outside the regime of applicability of our first-order $\lgb$ expansion; we do not expect that the decrease of the entropy around $\mu t = - 0.5$ would occur in a non-perturbative computation.
 
\noindent
{\bf 4. Discussion.---}In this Letter, we presented the first computation of holographic heavy ion collisions at finite coupling, which amounted to solving the collision of gravitational shock waves in Einstein-Gauss-Bonnet theory, perturbatively in $\lgb$. Interestingly, we found that the reduced coupling resulted in more energy on the light cone for narrow shocks, as well as less stopping for wider shocks, as measured by the reduced pile-up of energy.

Our work sheds light on finite coupling corrections to strongly coupled phenomenology of heavy ion collisions, in particular to its hydrodynamisation times, its rapidity profile and the entropy production (previously studied in \cite{vanderSchee:2015rta,Gubser:2008pc,Lin:2009pn,vanderSchee:2014qwa}). In \cite{vanderSchee:2015rta}, it was found that the rapidity profile at strong coupling needed to be wider by about 50\% to describe experimental data. Our results indeed indicate a wider initial rapidity profile, as is also suggested by the increased energy on the lightcone for narrow shock collisions. Nevertheless, quantitatively, this increase only amounts to approximately 20\%. Also, the increased viscosity leads to a reduced rapidity width and increased entropy production at later times, perhaps balancing each other at $\mu \tau$ = \{3, 6\} for \{narrow, wide\} shocks. However, in QCD, the quantitative size of these two effects can vary, especially as the shear viscosity of QCD in this regime is expected to decrease with decreasing temperature.

In this work, we considered only the simplest model for finite coupling corrections (perturbative curvature-squared in pure gravity), studied it to next-to-leading-order in (inverse) perturbative coupling corrections and thus made the first step towards phenomenologically more accurate holographic models. Studying the theory non-perturbatively in $\lgb$ would be much more challenging, as the nested structure of the characteristic formulation of general relativity would be lost. In future, it will also be important to compare our results to similar perturbative simulations in $\CN=4$ SYM theory. However, supported by the findings of ref. \cite{Grozdanov:2016vgg} which analysed the corrections to the linear spectrum from non-perturbative Gauss-Bonnet and perturbative $\alpha'^3$ terms in type IIB supergravity (dual to $\CN = 4$ SYM), we expect at least the qualitative behaviour of shock waves collisions in top-down constructions to remain similar to the results of this paper. It would also be interesting to study coupling corrections in non-conformal or charged theories (see \cite{Attems:2016tby, Casalderrey-Solana:2016xfq}), which could model a varying viscosity as well as non-trivial baryonic charge densities.

Finally, a full description of the initial stages of heavy ion collisions will likely involve insights from both weakly and strongly coupled physics, and a complete description will then require an interpolation between physics found at both weak and strong coupling, as e.g. discussed in \cite{Keegan:2015avk}. To make this interpolation, it will then be crucial to perform expansions both around $\lambda=0$ as well as around $\lambda=\infty$. For dynamical far-from-equilibrium collisions, the latter expansion will follow the procedure that we initiated in this work.

{\bf Acknowledgements.---}We thank Jorge Casalderrey Solana,  Andrei Starinets and especially Andrej Ficnar for helpful discussions. S. G. is supported by a VICI grant of the Netherlands Organization for Scientific Research (NWO), by the Netherlands Organization for Scientific Research/Ministry of Science and Education (NWO/OCW) and by the Foundation for Research into Fundamental Matter (FOM). W. S. is supported by the U.S. Department of Energy under grant Contract Number DE-SC0011090.

\bibliography{references}

\begin{thebibliography}{47}%
\makeatletter
\providecommand \@ifxundefined [1]{%
 \@ifx{#1\undefined}
}%
\providecommand \@ifnum [1]{%
 \ifnum #1\expandafter \@firstoftwo
 \else \expandafter \@secondoftwo
 \fi
}%
\providecommand \@ifx [1]{%
 \ifx #1\expandafter \@firstoftwo
 \else \expandafter \@secondoftwo
 \fi
}%
\providecommand \natexlab [1]{#1}%
\providecommand \enquote  [1]{``#1''}%
\providecommand \bibnamefont  [1]{#1}%
\providecommand \bibfnamefont [1]{#1}%
\providecommand \citenamefont [1]{#1}%
\providecommand \href@noop [0]{\@secondoftwo}%
\providecommand \href [0]{\begingroup \@sanitize@url \@href}%
\providecommand \@href[1]{\@@startlink{#1}\@@href}%
\providecommand \@@href[1]{\endgroup#1\@@endlink}%
\providecommand \@sanitize@url [0]{\catcode `\\12\catcode `\$12\catcode
  `\&12\catcode `\#12\catcode `\^12\catcode `\_12\catcode `\%12\relax}%
\providecommand \@@startlink[1]{}%
\providecommand \@@endlink[0]{}%
\providecommand \url  [0]{\begingroup\@sanitize@url \@url }%
\providecommand \@url [1]{\endgroup\@href {#1}{\urlprefix }}%
\providecommand \urlprefix  [0]{URL }%
\providecommand \Eprint [0]{\href }%
\providecommand \doibase [0]{http://dx.doi.org/}%
\providecommand \selectlanguage [0]{\@gobble}%
\providecommand \bibinfo  [0]{\@secondoftwo}%
\providecommand \bibfield  [0]{\@secondoftwo}%
\providecommand \translation [1]{[#1]}%
\providecommand \BibitemOpen [0]{}%
\providecommand \bibitemStop [0]{}%
\providecommand \bibitemNoStop [0]{.\EOS\space}%
\providecommand \EOS [0]{\spacefactor3000\relax}%
\providecommand \BibitemShut  [1]{\csname bibitem#1\endcsname}%
\let\auto@bib@innerbib\@empty
\bibitem [{\citenamefont {Chesler}\ and\ \citenamefont
  {Yaffe}(2011)}]{Chesler:2010bi}%
  \BibitemOpen
  \bibfield  {author} {\bibinfo {author} {\bibfnamefont {P.~M.}\ \bibnamefont
  {Chesler}}\ and\ \bibinfo {author} {\bibfnamefont {L.~G.}\ \bibnamefont
  {Yaffe}},\ }\href {\doibase 10.1103/PhysRevLett.106.021601} {\bibfield
  {journal} {\bibinfo  {journal} {Phys.Rev.Lett.}\ }\textbf {\bibinfo {volume}
  {106}},\ \bibinfo {pages} {021601} (\bibinfo {year} {2011})},\ \Eprint
  {http://arxiv.org/abs/1011.3562} {arXiv:1011.3562 [hep-th]} \BibitemShut
  {NoStop}%
\bibitem [{\citenamefont {Grumiller}\ and\ \citenamefont
  {Romatschke}(2008)}]{Grumiller:2008va}%
  \BibitemOpen
  \bibfield  {author} {\bibinfo {author} {\bibfnamefont {D.}~\bibnamefont
  {Grumiller}}\ and\ \bibinfo {author} {\bibfnamefont {P.}~\bibnamefont
  {Romatschke}},\ }\href {\doibase 10.1088/1126-6708/2008/08/027} {\bibfield
  {journal} {\bibinfo  {journal} {JHEP}\ }\textbf {\bibinfo {volume} {0808}},\
  \bibinfo {pages} {027} (\bibinfo {year} {2008})},\ \Eprint
  {http://arxiv.org/abs/0803.3226} {arXiv:0803.3226 [hep-th]} \BibitemShut
  {NoStop}%
\bibitem [{\citenamefont {Casalderrey-Solana}\ \emph
  {et~al.}(2013)\citenamefont {Casalderrey-Solana}, \citenamefont {Heller},
  \citenamefont {Mateos},\ and\ \citenamefont {van~der
  Schee}}]{Casalderrey-Solana:2013aba}%
  \BibitemOpen
  \bibfield  {author} {\bibinfo {author} {\bibfnamefont {J.}~\bibnamefont
  {Casalderrey-Solana}}, \bibinfo {author} {\bibfnamefont {M.~P.}\ \bibnamefont
  {Heller}}, \bibinfo {author} {\bibfnamefont {D.}~\bibnamefont {Mateos}}, \
  and\ \bibinfo {author} {\bibfnamefont {W.}~\bibnamefont {van~der Schee}},\
  }\href {\doibase 10.1103/PhysRevLett.111.181601} {\bibfield  {journal}
  {\bibinfo  {journal} {Phys. Rev. Lett. 111,}\ }\textbf {\bibinfo {volume}
  {181601}} (\bibinfo {year} {2013}),\ 10.1103/PhysRevLett.111.181601},\
  \Eprint {http://arxiv.org/abs/1305.4919} {arXiv:1305.4919 [hep-th]}
  \BibitemShut {NoStop}%
\bibitem [{\citenamefont {Casalderrey-Solana}\ \emph
  {et~al.}(2014{\natexlab{a}})\citenamefont {Casalderrey-Solana}, \citenamefont
  {Heller}, \citenamefont {Mateos},\ and\ \citenamefont {van~der
  Schee}}]{Casalderrey-Solana:2013sxa}%
  \BibitemOpen
  \bibfield  {author} {\bibinfo {author} {\bibfnamefont {J.}~\bibnamefont
  {Casalderrey-Solana}}, \bibinfo {author} {\bibfnamefont {M.~P.}\ \bibnamefont
  {Heller}}, \bibinfo {author} {\bibfnamefont {D.}~\bibnamefont {Mateos}}, \
  and\ \bibinfo {author} {\bibfnamefont {W.}~\bibnamefont {van~der Schee}},\
  }\href {\doibase 10.1103/PhysRevLett.112.221602} {\bibfield  {journal}
  {\bibinfo  {journal} {Phys. Rev. Lett.}\ }\textbf {\bibinfo {volume} {112}},\
  \bibinfo {pages} {221602} (\bibinfo {year} {2014}{\natexlab{a}})},\ \Eprint
  {http://arxiv.org/abs/1312.2956} {arXiv:1312.2956 [hep-th]} \BibitemShut
  {NoStop}%
\bibitem [{\citenamefont {Chesler}\ and\ \citenamefont
  {Yaffe}(2015)}]{Chesler:2015wra}%
  \BibitemOpen
  \bibfield  {author} {\bibinfo {author} {\bibfnamefont {P.~M.}\ \bibnamefont
  {Chesler}}\ and\ \bibinfo {author} {\bibfnamefont {L.~G.}\ \bibnamefont
  {Yaffe}},\ }\href {\doibase 10.1007/JHEP10(2015)070} {\bibfield  {journal}
  {\bibinfo  {journal} {JHEP}\ }\textbf {\bibinfo {volume} {10}},\ \bibinfo
  {pages} {070} (\bibinfo {year} {2015})},\ \Eprint
  {http://arxiv.org/abs/1501.04644} {arXiv:1501.04644 [hep-th]} \BibitemShut
  {NoStop}%
\bibitem [{\citenamefont {Casalderrey-Solana}\ \emph
  {et~al.}(2014{\natexlab{b}})\citenamefont {Casalderrey-Solana}, \citenamefont
  {Liu}, \citenamefont {Mateos}, \citenamefont {Rajagopal},\ and\ \citenamefont
  {Wiedemann}}]{jorge-book}%
  \BibitemOpen
  \bibfield  {author} {\bibinfo {author} {\bibfnamefont {J.}~\bibnamefont
  {Casalderrey-Solana}}, \bibinfo {author} {\bibfnamefont {H.}~\bibnamefont
  {Liu}}, \bibinfo {author} {\bibfnamefont {D.}~\bibnamefont {Mateos}},
  \bibinfo {author} {\bibfnamefont {K.}~\bibnamefont {Rajagopal}}, \ and\
  \bibinfo {author} {\bibfnamefont {U.~A.}\ \bibnamefont {Wiedemann}},\
  }\href@noop {} {\emph {\bibinfo {title} {{Gauge/String Duality, Hot QCD and
  Heavy Ion Collisions}}}}\ (\bibinfo  {publisher} {Cambridge University
  Press},\ \bibinfo {address} {Cambridge, UK},\ \bibinfo {year}
  {2014})\BibitemShut {NoStop}%
\bibitem [{\citenamefont {Chesler}\ and\ \citenamefont {van~der
  Schee}(2015)}]{Chesler:2015lsa}%
  \BibitemOpen
  \bibfield  {author} {\bibinfo {author} {\bibfnamefont {P.~M.}\ \bibnamefont
  {Chesler}}\ and\ \bibinfo {author} {\bibfnamefont {W.}~\bibnamefont {van~der
  Schee}},\ }\href {\doibase 10.1142/S0218301315300118} {\bibfield  {journal}
  {\bibinfo  {journal} {Int. J. Mod. Phys.}\ }\textbf {\bibinfo {volume}
  {E24}},\ \bibinfo {pages} {1530011} (\bibinfo {year} {2015})},\ \Eprint
  {http://arxiv.org/abs/1501.04952} {arXiv:1501.04952 [nucl-th]} \BibitemShut
  {NoStop}%
\bibitem [{\citenamefont {Heller}(2016)}]{Heller:2016gbp}%
  \BibitemOpen
  \bibfield  {author} {\bibinfo {author} {\bibfnamefont {M.~P.}\ \bibnamefont
  {Heller}}\ }(\bibinfo {year} {2016})\ \Eprint
  {http://arxiv.org/abs/1610.02023} {arXiv:1610.02023 [hep-th]} \BibitemShut
  {NoStop}%
\bibitem [{\citenamefont {Denef}\ \emph {et~al.}(2009)\citenamefont {Denef},
  \citenamefont {Hartnoll},\ and\ \citenamefont {Sachdev}}]{Denef:2009yy}%
  \BibitemOpen
  \bibfield  {author} {\bibinfo {author} {\bibfnamefont {F.}~\bibnamefont
  {Denef}}, \bibinfo {author} {\bibfnamefont {S.~A.}\ \bibnamefont {Hartnoll}},
  \ and\ \bibinfo {author} {\bibfnamefont {S.}~\bibnamefont {Sachdev}},\ }\href
  {\doibase 10.1103/PhysRevD.80.126016} {\bibfield  {journal} {\bibinfo
  {journal} {Phys. Rev.}\ }\textbf {\bibinfo {volume} {D80}},\ \bibinfo {pages}
  {126016} (\bibinfo {year} {2009})},\ \Eprint {http://arxiv.org/abs/0908.1788}
  {arXiv:0908.1788 [hep-th]} \BibitemShut {NoStop}%
\bibitem [{\citenamefont {Denef}\ \emph {et~al.}(2010)\citenamefont {Denef},
  \citenamefont {Hartnoll},\ and\ \citenamefont {Sachdev}}]{Denef:2009kn}%
  \BibitemOpen
  \bibfield  {author} {\bibinfo {author} {\bibfnamefont {F.}~\bibnamefont
  {Denef}}, \bibinfo {author} {\bibfnamefont {S.~A.}\ \bibnamefont {Hartnoll}},
  \ and\ \bibinfo {author} {\bibfnamefont {S.}~\bibnamefont {Sachdev}},\ }\href
  {\doibase 10.1088/0264-9381/27/12/125001} {\bibfield  {journal} {\bibinfo
  {journal} {Class. Quant. Grav.}\ }\textbf {\bibinfo {volume} {27}},\ \bibinfo
  {pages} {125001} (\bibinfo {year} {2010})},\ \Eprint
  {http://arxiv.org/abs/0908.2657} {arXiv:0908.2657 [hep-th]} \BibitemShut
  {NoStop}%
\bibitem [{\citenamefont {Arnold}\ \emph {et~al.}(2016)\citenamefont {Arnold},
  \citenamefont {Szepietowski},\ and\ \citenamefont {Vaman}}]{Arnold:2016dbb}%
  \BibitemOpen
  \bibfield  {author} {\bibinfo {author} {\bibfnamefont {P.}~\bibnamefont
  {Arnold}}, \bibinfo {author} {\bibfnamefont {P.}~\bibnamefont
  {Szepietowski}}, \ and\ \bibinfo {author} {\bibfnamefont {D.}~\bibnamefont
  {Vaman}},\ }\href@noop {} {\  (\bibinfo {year} {2016})},\ \Eprint
  {http://arxiv.org/abs/1603.08994} {arXiv:1603.08994 [hep-th]} \BibitemShut
  {NoStop}%
\bibitem [{\citenamefont {Gubser}\ \emph {et~al.}(1998)\citenamefont {Gubser},
  \citenamefont {Klebanov},\ and\ \citenamefont {Tseytlin}}]{Gubser:1998nz}%
  \BibitemOpen
  \bibfield  {author} {\bibinfo {author} {\bibfnamefont {S.~S.}\ \bibnamefont
  {Gubser}}, \bibinfo {author} {\bibfnamefont {I.~R.}\ \bibnamefont
  {Klebanov}}, \ and\ \bibinfo {author} {\bibfnamefont {A.~A.}\ \bibnamefont
  {Tseytlin}},\ }\href {\doibase 10.1016/S0550-3213(98)00514-8} {\bibfield
  {journal} {\bibinfo  {journal} {Nucl.Phys.}\ }\textbf {\bibinfo {volume}
  {B534}},\ \bibinfo {pages} {202} (\bibinfo {year} {1998})},\ \Eprint
  {http://arxiv.org/abs/hep-th/9805156} {arXiv:hep-th/9805156 [hep-th]}
  \BibitemShut {NoStop}%
\bibitem [{\citenamefont {Kats}\ and\ \citenamefont
  {Petrov}(2009)}]{Kats:2007mq}%
  \BibitemOpen
  \bibfield  {author} {\bibinfo {author} {\bibfnamefont {Y.}~\bibnamefont
  {Kats}}\ and\ \bibinfo {author} {\bibfnamefont {P.}~\bibnamefont {Petrov}},\
  }\href {\doibase 10.1088/1126-6708/2009/01/044} {\bibfield  {journal}
  {\bibinfo  {journal} {JHEP}\ }\textbf {\bibinfo {volume} {01}},\ \bibinfo
  {pages} {044} (\bibinfo {year} {2009})},\ \Eprint
  {http://arxiv.org/abs/0712.0743} {arXiv:0712.0743 [hep-th]} \BibitemShut
  {NoStop}%
\bibitem [{\citenamefont {Brigante}\ \emph
  {et~al.}(2008{\natexlab{a}})\citenamefont {Brigante}, \citenamefont {Liu},
  \citenamefont {Myers}, \citenamefont {Shenker},\ and\ \citenamefont
  {Yaida}}]{Brigante:2007nu}%
  \BibitemOpen
  \bibfield  {author} {\bibinfo {author} {\bibfnamefont {M.}~\bibnamefont
  {Brigante}}, \bibinfo {author} {\bibfnamefont {H.}~\bibnamefont {Liu}},
  \bibinfo {author} {\bibfnamefont {R.~C.}\ \bibnamefont {Myers}}, \bibinfo
  {author} {\bibfnamefont {S.}~\bibnamefont {Shenker}}, \ and\ \bibinfo
  {author} {\bibfnamefont {S.}~\bibnamefont {Yaida}},\ }\href {\doibase
  10.1103/PhysRevD.77.126006} {\bibfield  {journal} {\bibinfo  {journal} {Phys.
  Rev.}\ }\textbf {\bibinfo {volume} {D77}},\ \bibinfo {pages} {126006}
  (\bibinfo {year} {2008}{\natexlab{a}})},\ \Eprint
  {http://arxiv.org/abs/0712.0805} {arXiv:0712.0805 [hep-th]} \BibitemShut
  {NoStop}%
\bibitem [{\citenamefont {Brigante}\ \emph
  {et~al.}(2008{\natexlab{b}})\citenamefont {Brigante}, \citenamefont {Liu},
  \citenamefont {Myers}, \citenamefont {Shenker},\ and\ \citenamefont
  {Yaida}}]{Brigante:2008gz}%
  \BibitemOpen
  \bibfield  {author} {\bibinfo {author} {\bibfnamefont {M.}~\bibnamefont
  {Brigante}}, \bibinfo {author} {\bibfnamefont {H.}~\bibnamefont {Liu}},
  \bibinfo {author} {\bibfnamefont {R.~C.}\ \bibnamefont {Myers}}, \bibinfo
  {author} {\bibfnamefont {S.}~\bibnamefont {Shenker}}, \ and\ \bibinfo
  {author} {\bibfnamefont {S.}~\bibnamefont {Yaida}},\ }\href {\doibase
  10.1103/PhysRevLett.100.191601} {\bibfield  {journal} {\bibinfo  {journal}
  {Phys. Rev. Lett.}\ }\textbf {\bibinfo {volume} {100}},\ \bibinfo {pages}
  {191601} (\bibinfo {year} {2008}{\natexlab{b}})},\ \Eprint
  {http://arxiv.org/abs/0802.3318} {arXiv:0802.3318 [hep-th]} \BibitemShut
  {NoStop}%
\bibitem [{\citenamefont {Buchel}\ \emph {et~al.}(2008)\citenamefont {Buchel},
  \citenamefont {Myers}, \citenamefont {Paulos},\ and\ \citenamefont
  {Sinha}}]{Buchel:2008ae}%
  \BibitemOpen
  \bibfield  {author} {\bibinfo {author} {\bibfnamefont {A.}~\bibnamefont
  {Buchel}}, \bibinfo {author} {\bibfnamefont {R.~C.}\ \bibnamefont {Myers}},
  \bibinfo {author} {\bibfnamefont {M.~F.}\ \bibnamefont {Paulos}}, \ and\
  \bibinfo {author} {\bibfnamefont {A.}~\bibnamefont {Sinha}},\ }\href
  {\doibase 10.1016/j.physletb.2008.10.003} {\bibfield  {journal} {\bibinfo
  {journal} {Phys. Lett.}\ }\textbf {\bibinfo {volume} {B669}},\ \bibinfo
  {pages} {364} (\bibinfo {year} {2008})},\ \Eprint
  {http://arxiv.org/abs/0808.1837} {arXiv:0808.1837 [hep-th]} \BibitemShut
  {NoStop}%
\bibitem [{\citenamefont {Grozdanov}\ and\ \citenamefont
  {Starinets}(2015{\natexlab{a}})}]{Grozdanov:2014kva}%
  \BibitemOpen
  \bibfield  {author} {\bibinfo {author} {\bibfnamefont {S.}~\bibnamefont
  {Grozdanov}}\ and\ \bibinfo {author} {\bibfnamefont {A.~O.}\ \bibnamefont
  {Starinets}},\ }\href {\doibase 10.1007/JHEP03(2015)007} {\bibfield
  {journal} {\bibinfo  {journal} {JHEP}\ }\textbf {\bibinfo {volume} {03}},\
  \bibinfo {pages} {007} (\bibinfo {year} {2015}{\natexlab{a}})},\ \Eprint
  {http://arxiv.org/abs/1412.5685} {arXiv:1412.5685 [hep-th]} \BibitemShut
  {NoStop}%
\bibitem [{\citenamefont {Grozdanov}\ and\ \citenamefont
  {Starinets}(2015{\natexlab{b}})}]{Grozdanov:2015asa}%
  \BibitemOpen
  \bibfield  {author} {\bibinfo {author} {\bibfnamefont {S.}~\bibnamefont
  {Grozdanov}}\ and\ \bibinfo {author} {\bibfnamefont {A.~O.}\ \bibnamefont
  {Starinets}},\ }\href {\doibase 10.1007/s11232-015-0245-7} {\bibfield
  {journal} {\bibinfo  {journal} {Theor. Math. Phys.}\ }\textbf {\bibinfo
  {volume} {182}},\ \bibinfo {pages} {61} (\bibinfo {year}
  {2015}{\natexlab{b}})},\ \bibinfo {note} {[Teor. Mat.
  Fiz.182,no.1,76(2014)]}\BibitemShut {NoStop}%
\bibitem [{\citenamefont {Stricker}(2014)}]{Stricker:2013lma}%
  \BibitemOpen
  \bibfield  {author} {\bibinfo {author} {\bibfnamefont {S.~A.}\ \bibnamefont
  {Stricker}},\ }\href {\doibase 10.1140/epjc/s10052-014-2727-4} {\bibfield
  {journal} {\bibinfo  {journal} {Eur. Phys. J.}\ }\textbf {\bibinfo {volume}
  {C74}},\ \bibinfo {pages} {2727} (\bibinfo {year} {2014})},\ \Eprint
  {http://arxiv.org/abs/1307.2736} {arXiv:1307.2736 [hep-th]} \BibitemShut
  {NoStop}%
\bibitem [{\citenamefont {Waeber}\ \emph {et~al.}(2015)\citenamefont {Waeber},
  \citenamefont {Schaefer}, \citenamefont {Vuorinen},\ and\ \citenamefont
  {Yaffe}}]{Waeber:2015oka}%
  \BibitemOpen
  \bibfield  {author} {\bibinfo {author} {\bibfnamefont {S.}~\bibnamefont
  {Waeber}}, \bibinfo {author} {\bibfnamefont {A.}~\bibnamefont {Schaefer}},
  \bibinfo {author} {\bibfnamefont {A.}~\bibnamefont {Vuorinen}}, \ and\
  \bibinfo {author} {\bibfnamefont {L.~G.}\ \bibnamefont {Yaffe}},\ }\href
  {\doibase 10.1007/JHEP11(2015)087} {\bibfield  {journal} {\bibinfo  {journal}
  {JHEP}\ }\textbf {\bibinfo {volume} {11}},\ \bibinfo {pages} {087} (\bibinfo
  {year} {2015})},\ \Eprint {http://arxiv.org/abs/1509.02983} {arXiv:1509.02983
  [hep-th]} \BibitemShut {NoStop}%
\bibitem [{\citenamefont {Grozdanov}\ \emph {et~al.}(2016)\citenamefont
  {Grozdanov}, \citenamefont {Kaplis},\ and\ \citenamefont
  {Starinets}}]{Grozdanov:2016vgg}%
  \BibitemOpen
  \bibfield  {author} {\bibinfo {author} {\bibfnamefont {S.}~\bibnamefont
  {Grozdanov}}, \bibinfo {author} {\bibfnamefont {N.}~\bibnamefont {Kaplis}}, \
  and\ \bibinfo {author} {\bibfnamefont {A.~O.}\ \bibnamefont {Starinets}},\
  }\href {\doibase 10.1007/JHEP07(2016)151} {\bibfield  {journal} {\bibinfo
  {journal} {JHEP}\ }\textbf {\bibinfo {volume} {07}},\ \bibinfo {pages} {151}
  (\bibinfo {year} {2016})},\ \Eprint {http://arxiv.org/abs/1605.02173}
  {arXiv:1605.02173 [hep-th]} \BibitemShut {NoStop}%
\bibitem [{\citenamefont {Andrade}\ \emph
  {et~al.}(2016{\natexlab{a}})\citenamefont {Andrade}, \citenamefont
  {Casalderrey-Solana},\ and\ \citenamefont {Ficnar}}]{Andrade:2016aaa}%
  \BibitemOpen
  \bibfield  {author} {\bibinfo {author} {\bibfnamefont {T.}~\bibnamefont
  {Andrade}}, \bibinfo {author} {\bibfnamefont {J.}~\bibnamefont
  {Casalderrey-Solana}}, \ and\ \bibinfo {author} {\bibfnamefont
  {A.}~\bibnamefont {Ficnar}},\ }\href@noop {} {\  (\bibinfo {year}
  {2016}{\natexlab{a}})}\BibitemShut {NoStop}%
\bibitem [{\citenamefont {Callan}\ \emph {et~al.}(1987)\citenamefont {Callan},
  \citenamefont {Lovelace}, \citenamefont {Nappi},\ and\ \citenamefont
  {Yost}}]{Callan:1986bc}%
  \BibitemOpen
  \bibfield  {author} {\bibinfo {author} {\bibfnamefont {C.~G.}\ \bibnamefont
  {Callan}, \bibfnamefont {Jr.}}, \bibinfo {author} {\bibfnamefont
  {C.}~\bibnamefont {Lovelace}}, \bibinfo {author} {\bibfnamefont {C.~R.}\
  \bibnamefont {Nappi}}, \ and\ \bibinfo {author} {\bibfnamefont {S.~A.}\
  \bibnamefont {Yost}},\ }\href {\doibase 10.1016/0550-3213(87)90227-6}
  {\bibfield  {journal} {\bibinfo  {journal} {Nucl. Phys.}\ }\textbf {\bibinfo
  {volume} {B288}},\ \bibinfo {pages} {525} (\bibinfo {year}
  {1987})}\BibitemShut {NoStop}%
\bibitem [{\citenamefont {Grisaru}\ and\ \citenamefont
  {Zanon}(1986)}]{Grisaru:1986vi}%
  \BibitemOpen
  \bibfield  {author} {\bibinfo {author} {\bibfnamefont {M.~T.}\ \bibnamefont
  {Grisaru}}\ and\ \bibinfo {author} {\bibfnamefont {D.}~\bibnamefont
  {Zanon}},\ }\href {\doibase 10.1016/0370-2693(86)90765-3} {\bibfield
  {journal} {\bibinfo  {journal} {Phys. Lett.}\ }\textbf {\bibinfo {volume}
  {B177}},\ \bibinfo {pages} {347} (\bibinfo {year} {1986})}\BibitemShut
  {NoStop}%
\bibitem [{\citenamefont {Gross}\ and\ \citenamefont
  {Sloan}(1987)}]{Gross:1986mw}%
  \BibitemOpen
  \bibfield  {author} {\bibinfo {author} {\bibfnamefont {D.~J.}\ \bibnamefont
  {Gross}}\ and\ \bibinfo {author} {\bibfnamefont {J.~H.}\ \bibnamefont
  {Sloan}},\ }\href {\doibase 10.1016/0550-3213(87)90465-2} {\bibfield
  {journal} {\bibinfo  {journal} {Nucl. Phys.}\ }\textbf {\bibinfo {volume}
  {B291}},\ \bibinfo {pages} {41} (\bibinfo {year} {1987})}\BibitemShut
  {NoStop}%
\bibitem [{\citenamefont {Gross}\ and\ \citenamefont
  {Witten}(1986)}]{Gross:1986iv}%
  \BibitemOpen
  \bibfield  {author} {\bibinfo {author} {\bibfnamefont {D.~J.}\ \bibnamefont
  {Gross}}\ and\ \bibinfo {author} {\bibfnamefont {E.}~\bibnamefont {Witten}},\
  }\href {\doibase 10.1016/0550-3213(86)90429-3} {\bibfield  {journal}
  {\bibinfo  {journal} {Nucl. Phys.}\ }\textbf {\bibinfo {volume} {B277}},\
  \bibinfo {pages} {1} (\bibinfo {year} {1986})}\BibitemShut {NoStop}%
\bibitem [{\citenamefont {Freeman}\ \emph {et~al.}(1986)\citenamefont
  {Freeman}, \citenamefont {Pope}, \citenamefont {Sohnius},\ and\ \citenamefont
  {Stelle}}]{Freeman:1986zh}%
  \BibitemOpen
  \bibfield  {author} {\bibinfo {author} {\bibfnamefont {M.~D.}\ \bibnamefont
  {Freeman}}, \bibinfo {author} {\bibfnamefont {C.~N.}\ \bibnamefont {Pope}},
  \bibinfo {author} {\bibfnamefont {M.~F.}\ \bibnamefont {Sohnius}}, \ and\
  \bibinfo {author} {\bibfnamefont {K.~S.}\ \bibnamefont {Stelle}},\ }\href
  {\doibase 10.1016/0370-2693(86)91495-4} {\bibfield  {journal} {\bibinfo
  {journal} {Phys. Lett.}\ }\textbf {\bibinfo {volume} {B178}},\ \bibinfo
  {pages} {199} (\bibinfo {year} {1986})}\BibitemShut {NoStop}%
\bibitem [{Note1()}]{Note1}%
  \BibitemOpen
  \bibinfo {note} {Any perturbative $R^2$ theory can be conveniently
  transformed into the Einstein-Gauss-Bonnet theory (with second-order
  equations of motion). For details, see \cite
  {Brigante:2007nu,Grozdanov:2014kva,Grozdanov:2016vgg}.}\BibitemShut {Stop}%
\bibitem [{Note2()}]{Note2}%
  \BibitemOpen
  \bibinfo {note} {Recently, \cite {Camanho:2014apa} argued that \protect
  \textup {\hbox {\mathsurround \z@ \protect \normalfont (\ignorespaces \ref
  {GBaction}\unskip \@@italiccorr )}} violates causality unless $|\lambda
  _{\scriptscriptstyle GB}| / L^2 \ll 1$ (see however \cite
  {Papallo:2015rna,Andrade:2016yzc}). Since we work perturbatively in $\lambda
  _{\scriptscriptstyle GB}$, such restrictions should not affect our
  findings.}\BibitemShut {Stop}%
\bibitem [{Note3()}]{Note3}%
  \BibitemOpen
  \bibinfo {note} {An advantage of this choice is that the non-normalisable
  mode of the metric does not receive corrections.}\BibitemShut {Stop}%
\bibitem [{\citenamefont {Grozdanov}\ and\ \citenamefont
  {Starinets}(2017)}]{Grozdanov:2016fkt}%
  \BibitemOpen
  \bibfield  {author} {\bibinfo {author} {\bibfnamefont {S.}~\bibnamefont
  {Grozdanov}}\ and\ \bibinfo {author} {\bibfnamefont {A.~O.}\ \bibnamefont
  {Starinets}},\ }\href {\doibase 10.1007/JHEP03(2017)166} {\bibfield
  {journal} {\bibinfo  {journal} {JHEP}\ }\textbf {\bibinfo {volume} {03}},\
  \bibinfo {pages} {166} (\bibinfo {year} {2017})},\ \Eprint
  {http://arxiv.org/abs/1611.07053} {arXiv:1611.07053 [hep-th]} \BibitemShut
  {NoStop}%
\bibitem [{\citenamefont {Bondi}(1960)}]{Bondi1960Gravitational}%
  \BibitemOpen
  \bibfield  {author} {\bibinfo {author} {\bibfnamefont {H.}~\bibnamefont
  {Bondi}},\ }\href {http://dx.doi.org/10.1038/186535a0} {\bibfield  {journal}
  {\bibinfo  {journal} {Nature}\ }\textbf {\bibinfo {volume} {186}},\ \bibinfo
  {pages} {535+} (\bibinfo {year} {1960})}\BibitemShut {NoStop}%
\bibitem [{\citenamefont {Brihaye}\ and\ \citenamefont
  {Radu}(2008)}]{Brihaye:2008kh}%
  \BibitemOpen
  \bibfield  {author} {\bibinfo {author} {\bibfnamefont {Y.}~\bibnamefont
  {Brihaye}}\ and\ \bibinfo {author} {\bibfnamefont {E.}~\bibnamefont {Radu}},\
  }\href {\doibase 10.1016/j.physletb.2008.02.005} {\bibfield  {journal}
  {\bibinfo  {journal} {Phys. Lett.}\ }\textbf {\bibinfo {volume} {B661}},\
  \bibinfo {pages} {167} (\bibinfo {year} {2008})},\ \Eprint
  {http://arxiv.org/abs/0801.1021} {arXiv:0801.1021 [hep-th]} \BibitemShut
  {NoStop}%
\bibitem [{\citenamefont {Janik}\ and\ \citenamefont
  {Peschanski}(2006)}]{Janik:2005zt}%
  \BibitemOpen
  \bibfield  {author} {\bibinfo {author} {\bibfnamefont {R.~A.}\ \bibnamefont
  {Janik}}\ and\ \bibinfo {author} {\bibfnamefont {R.~B.}\ \bibnamefont
  {Peschanski}},\ }\href {\doibase 10.1103/PhysRevD.73.045013} {\bibfield
  {journal} {\bibinfo  {journal} {Phys.Rev.}\ }\textbf {\bibinfo {volume}
  {D73}},\ \bibinfo {pages} {045013} (\bibinfo {year} {2006})},\ \Eprint
  {http://arxiv.org/abs/hep-th/0512162} {arXiv:hep-th/0512162 [hep-th]}
  \BibitemShut {NoStop}%
\bibitem [{\citenamefont {Chesler}\ and\ \citenamefont
  {Yaffe}(2014)}]{Chesler:2013lia}%
  \BibitemOpen
  \bibfield  {author} {\bibinfo {author} {\bibfnamefont {P.~M.}\ \bibnamefont
  {Chesler}}\ and\ \bibinfo {author} {\bibfnamefont {L.~G.}\ \bibnamefont
  {Yaffe}},\ }\href {\doibase 10.1007/JHEP07(2014)086} {\bibfield  {journal}
  {\bibinfo  {journal} {JHEP}\ }\textbf {\bibinfo {volume} {07}},\ \bibinfo
  {pages} {086} (\bibinfo {year} {2014})},\ \Eprint
  {http://arxiv.org/abs/1309.1439} {arXiv:1309.1439 [hep-th]} \BibitemShut
  {NoStop}%
\bibitem [{\citenamefont {van~der Schee}(2014)}]{vanderSchee:2014qwa}%
  \BibitemOpen
  \bibfield  {author} {\bibinfo {author} {\bibfnamefont {W.}~\bibnamefont
  {van~der Schee}},\ }\emph {\bibinfo {title} {{Gravitational collisions and
  the quark-gluon plasma}}},\ \href@noop {} {Ph.D. thesis},\ \bibinfo  {school}
  {Utrecht University} (\bibinfo {year} {2014}),\ \Eprint
  {http://arxiv.org/abs/1407.1849} {arXiv:1407.1849 [hep-th]} \BibitemShut
  {NoStop}%
\bibitem [{Note4()}]{Note4}%
  \BibitemOpen
  \bibinfo {note} {For numerical stability, our simulations include a regulator
  energy density of less than 1\% of the peak energy density. Further technical
  specifications, including the \protect \emph {Mathematica} code used for the
  simulations, can be found on \protect \href
  {http://sites.google.com/site/wilkevanderschee/}{sites.google.com/site/wilkevanderschee/}.}\BibitemShut
  {Stop}%
\bibitem [{Note5()}]{Note5}%
  \BibitemOpen
  \bibinfo {note} {The calculation to second order in $\lambda
  _{\scriptscriptstyle GB}$ is entirely analogous to the first-order
  calculation described above. We expand $A = A_0 + \lambda
  _{\scriptscriptstyle GB}A_1 +\lambda _{\scriptscriptstyle GB}^2 A_2$ and set
  $L =1+\lambda _{\scriptscriptstyle GB}/2 + 7 \lambda _{\scriptscriptstyle
  GB}^2 / 8$, which gives $\protect \mathcal {E} = -\protect \frac {3}{4} \left
  (a_{4,0} + \lambda _{\scriptscriptstyle GB}(a_{4,1}-2 a_{4,0}) +\lambda
  _{\scriptscriptstyle GB}^2 ( a_{4,2} - 2 a_{4,1} - 2 a_{4,0} ) \right )$. Our
  initial conditions thus require setting $a_{4,2} = 2 a_{4,1} + 2
  a_{4,0}$.}\BibitemShut {Stop}%
\bibitem [{\citenamefont {van~der Schee}\ and\ \citenamefont
  {Schenke}(2015)}]{vanderSchee:2015rta}%
  \BibitemOpen
  \bibfield  {author} {\bibinfo {author} {\bibfnamefont {W.}~\bibnamefont
  {van~der Schee}}\ and\ \bibinfo {author} {\bibfnamefont {B.}~\bibnamefont
  {Schenke}},\ }\href {\doibase 10.1103/PhysRevC.92.064907} {\bibfield
  {journal} {\bibinfo  {journal} {Phys. Rev.}\ }\textbf {\bibinfo {volume}
  {C92}},\ \bibinfo {pages} {064907} (\bibinfo {year} {2015})},\ \Eprint
  {http://arxiv.org/abs/1507.08195} {arXiv:1507.08195 [nucl-th]} \BibitemShut
  {NoStop}%
\bibitem [{\citenamefont {Gubser}\ \emph {et~al.}(2008)\citenamefont {Gubser},
  \citenamefont {Pufu},\ and\ \citenamefont {Yarom}}]{Gubser:2008pc}%
  \BibitemOpen
  \bibfield  {author} {\bibinfo {author} {\bibfnamefont {S.~S.}\ \bibnamefont
  {Gubser}}, \bibinfo {author} {\bibfnamefont {S.~S.}\ \bibnamefont {Pufu}}, \
  and\ \bibinfo {author} {\bibfnamefont {A.}~\bibnamefont {Yarom}},\ }\href
  {\doibase 10.1103/PhysRevD.78.066014} {\bibfield  {journal} {\bibinfo
  {journal} {Phys.Rev.}\ }\textbf {\bibinfo {volume} {D78}},\ \bibinfo {pages}
  {066014} (\bibinfo {year} {2008})},\ \Eprint {http://arxiv.org/abs/0805.1551}
  {arXiv:0805.1551 [hep-th]} \BibitemShut {NoStop}%
\bibitem [{\citenamefont {Lin}\ and\ \citenamefont
  {Shuryak}(2009)}]{Lin:2009pn}%
  \BibitemOpen
  \bibfield  {author} {\bibinfo {author} {\bibfnamefont {S.}~\bibnamefont
  {Lin}}\ and\ \bibinfo {author} {\bibfnamefont {E.}~\bibnamefont {Shuryak}},\
  }\href {\doibase 10.1103/PhysRevD.79.124015} {\bibfield  {journal} {\bibinfo
  {journal} {Phys.Rev.}\ }\textbf {\bibinfo {volume} {D79}},\ \bibinfo {pages}
  {124015} (\bibinfo {year} {2009})},\ \Eprint {http://arxiv.org/abs/0902.1508}
  {arXiv:0902.1508 [hep-th]} \BibitemShut {NoStop}%
\bibitem [{\citenamefont {Attems}\ \emph {et~al.}(2016)\citenamefont {Attems},
  \citenamefont {Casalderrey-Solana}, \citenamefont {Mateos}, \citenamefont
  {Santos-Oliván}, \citenamefont {Sopuerta}, \citenamefont {Triana},\ and\
  \citenamefont {Zilhão}}]{Attems:2016tby}%
  \BibitemOpen
  \bibfield  {author} {\bibinfo {author} {\bibfnamefont {M.}~\bibnamefont
  {Attems}}, \bibinfo {author} {\bibfnamefont {J.}~\bibnamefont
  {Casalderrey-Solana}}, \bibinfo {author} {\bibfnamefont {D.}~\bibnamefont
  {Mateos}}, \bibinfo {author} {\bibfnamefont {D.}~\bibnamefont
  {Santos-Oliván}}, \bibinfo {author} {\bibfnamefont {C.~F.}\ \bibnamefont
  {Sopuerta}}, \bibinfo {author} {\bibfnamefont {M.}~\bibnamefont {Triana}}, \
  and\ \bibinfo {author} {\bibfnamefont {M.}~\bibnamefont {Zilhão}},\
  }\href@noop {} {\  (\bibinfo {year} {2016})},\ \Eprint
  {http://arxiv.org/abs/1604.06439} {arXiv:1604.06439 [hep-th]} \BibitemShut
  {NoStop}%
\bibitem [{\citenamefont {Casalderrey-Solana}\ \emph
  {et~al.}(2016)\citenamefont {Casalderrey-Solana}, \citenamefont {Mateos},
  \citenamefont {van~der Schee},\ and\ \citenamefont
  {Triana}}]{Casalderrey-Solana:2016xfq}%
  \BibitemOpen
  \bibfield  {author} {\bibinfo {author} {\bibfnamefont {J.}~\bibnamefont
  {Casalderrey-Solana}}, \bibinfo {author} {\bibfnamefont {D.}~\bibnamefont
  {Mateos}}, \bibinfo {author} {\bibfnamefont {W.}~\bibnamefont {van~der
  Schee}}, \ and\ \bibinfo {author} {\bibfnamefont {M.}~\bibnamefont
  {Triana}},\ }\href {\doibase 10.1007/JHEP09(2016)108} {\bibfield  {journal}
  {\bibinfo  {journal} {JHEP}\ }\textbf {\bibinfo {volume} {09}},\ \bibinfo
  {pages} {108} (\bibinfo {year} {2016})},\ \Eprint
  {http://arxiv.org/abs/1607.05273} {arXiv:1607.05273 [hep-th]} \BibitemShut
  {NoStop}%
\bibitem [{\citenamefont {Keegan}\ \emph {et~al.}(2016)\citenamefont {Keegan},
  \citenamefont {Kurkela}, \citenamefont {Romatschke}, \citenamefont {van~der
  Schee},\ and\ \citenamefont {Zhu}}]{Keegan:2015avk}%
  \BibitemOpen
  \bibfield  {author} {\bibinfo {author} {\bibfnamefont {L.}~\bibnamefont
  {Keegan}}, \bibinfo {author} {\bibfnamefont {A.}~\bibnamefont {Kurkela}},
  \bibinfo {author} {\bibfnamefont {P.}~\bibnamefont {Romatschke}}, \bibinfo
  {author} {\bibfnamefont {W.}~\bibnamefont {van~der Schee}}, \ and\ \bibinfo
  {author} {\bibfnamefont {Y.}~\bibnamefont {Zhu}},\ }\href {\doibase
  10.1007/JHEP04(2016)031} {\bibfield  {journal} {\bibinfo  {journal} {JHEP}\
  }\textbf {\bibinfo {volume} {04}},\ \bibinfo {pages} {031} (\bibinfo {year}
  {2016})},\ \Eprint {http://arxiv.org/abs/1512.05347} {arXiv:1512.05347
  [hep-th]} \BibitemShut {NoStop}%
\bibitem [{\citenamefont {Camanho}\ \emph {et~al.}(2016)\citenamefont
  {Camanho}, \citenamefont {Edelstein}, \citenamefont {Maldacena},\ and\
  \citenamefont {Zhiboedov}}]{Camanho:2014apa}%
  \BibitemOpen
  \bibfield  {author} {\bibinfo {author} {\bibfnamefont {X.~O.}\ \bibnamefont
  {Camanho}}, \bibinfo {author} {\bibfnamefont {J.~D.}\ \bibnamefont
  {Edelstein}}, \bibinfo {author} {\bibfnamefont {J.}~\bibnamefont
  {Maldacena}}, \ and\ \bibinfo {author} {\bibfnamefont {A.}~\bibnamefont
  {Zhiboedov}},\ }\href {\doibase 10.1007/JHEP02(2016)020} {\bibfield
  {journal} {\bibinfo  {journal} {JHEP}\ }\textbf {\bibinfo {volume} {02}},\
  \bibinfo {pages} {020} (\bibinfo {year} {2016})},\ \Eprint
  {http://arxiv.org/abs/1407.5597} {arXiv:1407.5597 [hep-th]} \BibitemShut
  {NoStop}%
\bibitem [{\citenamefont {Papallo}\ and\ \citenamefont
  {Reall}(2015)}]{Papallo:2015rna}%
  \BibitemOpen
  \bibfield  {author} {\bibinfo {author} {\bibfnamefont {G.}~\bibnamefont
  {Papallo}}\ and\ \bibinfo {author} {\bibfnamefont {H.~S.}\ \bibnamefont
  {Reall}},\ }\href {\doibase 10.1007/JHEP11(2015)109} {\bibfield  {journal}
  {\bibinfo  {journal} {JHEP}\ }\textbf {\bibinfo {volume} {11}},\ \bibinfo
  {pages} {109} (\bibinfo {year} {2015})},\ \Eprint
  {http://arxiv.org/abs/1508.05303} {arXiv:1508.05303 [gr-qc]} \BibitemShut
  {NoStop}%
\bibitem [{\citenamefont {Andrade}\ \emph
  {et~al.}(2016{\natexlab{b}})\citenamefont {Andrade}, \citenamefont
  {Caceres},\ and\ \citenamefont {Keeler}}]{Andrade:2016yzc}%
  \BibitemOpen
  \bibfield  {author} {\bibinfo {author} {\bibfnamefont {T.}~\bibnamefont
  {Andrade}}, \bibinfo {author} {\bibfnamefont {E.}~\bibnamefont {Caceres}}, \
  and\ \bibinfo {author} {\bibfnamefont {C.}~\bibnamefont {Keeler}},\
  }\href@noop {} {\  (\bibinfo {year} {2016}{\natexlab{b}})},\ \Eprint
  {http://arxiv.org/abs/1610.06078} {arXiv:1610.06078 [hep-th]} \BibitemShut
  {NoStop}%
\end{thebibliography}%
\end{document}